\documentclass[review]{elsarticle}

\usepackage{lineno,hyperref}
\modulolinenumbers[5]

\usepackage{amsmath}
\usepackage[utf8]{inputenc}
\usepackage[english]{babel}
\usepackage{mathtools}
\usepackage{amsthm}
\usepackage{amssymb}
\usepackage{multirow}

\newcommand{\bt}{\begin{table}[!h]}
\newcommand{\et}{\end{table}}

\renewcommand{\arraystretch}{0.6}
\makeatletter
\def\ps@pprintTitle{%
\let\@oddhead\@empty
\let\@evenhead\@empty
\def\@oddfoot{\centerline{\thepage}}%
\let\@evenfoot\@oddfoot}
\makeatother

\begin{document}

\begin{frontmatter}

\title{Power Generalized DUS Transformation of Exponential Distribution}

\author{ Beenu Thomas and Chacko V. M.}
\address{Department of Statistics}


\author[mysecondaryaddress]{St. Thomas' College (Autonomous)Thrissur, Kerala, India}
\ead{chackovm@gmail.com}

\begin{abstract}DUS transformation of lifetime distributions
received attention by engineers and researchers in recent years. The
present study introduces a new class of distribution using
exponentiation of DUS transformation. A new distribution using the
Exponential distribution as the baseline distribution in this
transformation is proposed. The statistical properties of  the
proposed distribution have been examined and the parameter
estimation is done using the method of maximum likelihood. The
fitness of the proposed model is established using real data
analysis.
\end{abstract}

\begin{keyword}
 DUS transformation \sep Exponential Distribution \sep Failure Rate \sep Moments\sep Maximum likelihood estimator

\end{keyword}

\end{frontmatter}

\section{Introduction}
\par Modeling and analysis of lifetime distribution have been extensively used in many fields of science like engineering and
statistics. Fitting of appropriate distribution is essential for a
proper data anlysis.
 Different methods are available that propose  new classes of distributions using existing distributions, see  Gupta et al.{\cite{GGG1998}}, Nadarajah and
Kotz{\cite{NK2006}}, Cordeiro and Castro {\cite{cc2011}}, Cordeiro
et al.{\cite{cod2013}}, Kumar et al.{\cite{SDUS2015}}, etc. Kumar et
al.{\cite{DUS2015}} proposed a method called DUS transformation to
obtain a new parsimonious class of distribution. Recently, Maurya et
al.{\cite{MKSS2016}} proposed a generalization to DUS transformation
to make it more flexible. Deepthi and Chacko{\cite{DC2020}}
introduced DUS Lomax distribution. Gauthami and
Chacko{\cite{GC2021}} introduced DUS Inverse Weibul distribution.
But the existing approach is not appropriate for some data. A search
for distributions with  better fit is quite essential for data
analysis in statistics and reliability engineering.
\par
The current research work aims to introduce a new class of
distribution using the exponentiation of DUS transformation, called
Power generalized DUS (PGDUS) tranformation. The new PGDUS
transformed distribution can be obtained as follows: Let X be a
random variable with baseline cumulative distribution function (cdf)
F(x) and corresponding probability density function (pdf) f(x). Then
the cdf of the proposed PGDUS distribution is defined as,
\begin{equation}\label{eq1}
G(x)=\left(\frac{e^{F(x)}-1}{e-1}\right)^\theta,\theta >0, \,x > 0.
\end{equation}
and the corresponding pdf is,
\begin{equation}
g(x)=\frac{\theta}{(e-1)^\theta}(e^{F(x)}-1)^{\theta-1}e^{F(x)}f(x),\theta>0,
\,x > 0.
\end{equation}
The associated survival function is,
$$S(x)=1-\big(\frac{e^{F(x)}-1}{e-1}\big)^\theta,\theta>0,\,x > 0.$$
The failure rate function is,
$$h(x)=\frac{\theta f(x)e^{F(x)}(e^{F(x)}-1)^{\theta-1}}{(e-1)^\theta-(e^{F(x)}-1)^\theta},\theta>0,\,x > 0.$$

Application of the new transformation to the existing distributions
has to be investigated.  Using Exponential distribution as baseline
distribution, Power Generalized DUS Exponential (PGDUSE)
distribution is proposed, in this paper. It has to be studied in
detail.

\par The rest of the paper is organized as follows. In Section 2,
the PGDUSE is proposed. Sections 3 discussed the statistical
properties of the proposed distribution. In Section 4, the maximum
likelihood estimation procedure  is applied for estimation of
parameters. Real data set is analyzed in Section 5.  Concluding
remarks are given in Section 6.

\section{Power Generalized DUS Exponential Distribution}
\begin{figure}[h!]
  \centering
  \includegraphics[width=4in]{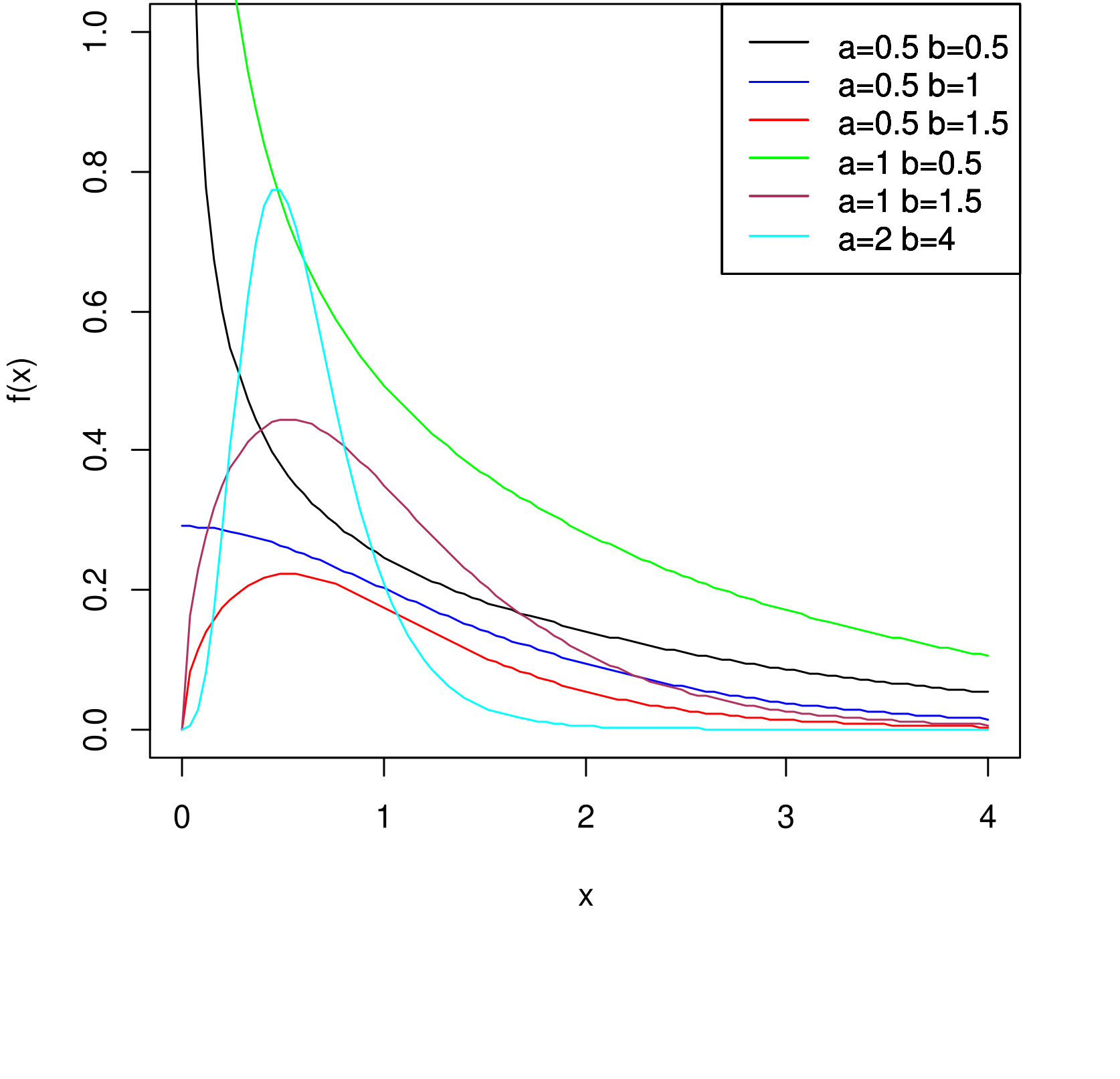}\\
  \caption{Density plot}\label{fig1}
\end{figure}

Here, Power Generalized DUS transformation to the baseline
distribution, Exponential distribution, is considered.  Consider the
Exponential distribution with parameter $\lambda$ as the baseline
distribution. Invoking the PGDUS transformation given in
equation(1), the cdf of the  PGDUSE distribution is obtained as
\begin{equation}
G(x)=\bigg(\frac{e^{1-e^{-\lambda x}}-1}{e-1}\bigg)^\theta,\lambda > 0, \theta > 0, x > 0.
\end{equation}
and the corresponding pdf is given by,
\begin{equation}
g(x)=\frac{\theta\lambda e^{1-\lambda x-e^{-\lambda x}}(e^{1-e^{-\lambda x}}-1)^{\theta-1}}{(e-1)^\theta},\lambda > 0, \theta > 0, x > 0.
\end{equation}
Then, the associated failure rate function is,
\begin{equation}
h(x)=\frac{\theta\lambda e^{1-\lambda x-e^{-\lambda x}}(e^{1-e^{-\lambda x}}-1)^{\theta-1}}{(e-1)^\theta-(e^{1-e^{-\lambda x}}-1)^\theta},\lambda > 0, \theta > 0, x > 0.
\end{equation}
We denote $PGDUSE(\lambda,\theta)$ for PGDUSE  distribution with
parameters $\lambda $ and $ \theta$.
 Figure \ref{fig1} shows that the density function of PGDUSE distribution is likely to be unimodal.
\begin{figure}[h!]
  \centering
  \includegraphics[width=4in]{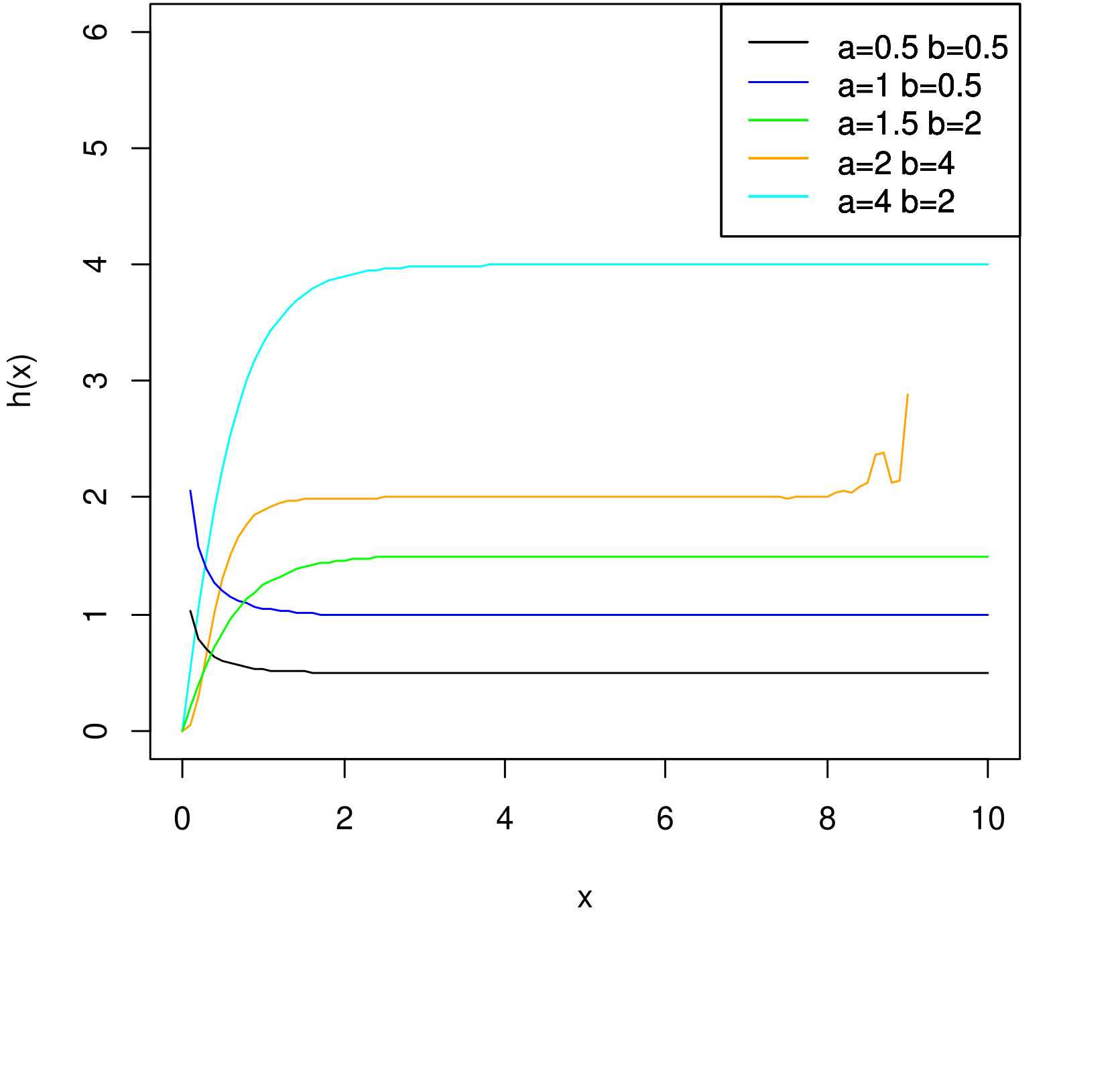}\\
  \caption{Failure rate plot}\label{fig2}
\end{figure}

\section{\textbf{Statistical Properties}}
For a distribution, the statistical properties are inevitable.  In
this section, a few statistical properties like moments, moment
generating function, characteristic function, cumulant generating
function, quantile function, order statistics, and entropy of the
proposed PGDUSE distribution are derived.

\subsection{\textbf{Moments}}
The rth raw moment of the $PGDUS-E(\lambda,\theta)$ distribution is given by
\begin{align*}
\mu'_r&=E(X^r)\\
~\\
&=\int_0^\infty x^r\frac{\theta\lambda}{(e-1)^\theta}e^{1-\lambda x-e^{-\lambda x}}(e^{1-e^{-\lambda x}}-1)dx\\
~\\
&=\frac{\theta\lambda}{(e-1)^\theta}\int_0^\infty x^r e^{2-\lambda x-2e^{-\lambda x}}dx-\frac{\theta\lambda}{(e-1)^\theta}\int_0^\infty x^r e^{1-\lambda x-e^{-\lambda x}}dx\\
~\\
&=\frac{\theta\lambda e^2}{(e-1)^\theta}\sum_{m=0}^{\infty}\frac{(-1)^m 2^m r!}{m!(1+m)^{r+1}\lambda^{r+1}}-\frac{\theta\lambda e}{(e-1)^\theta}\sum_{m=0}^{\infty}\frac{(-1)^m r!}{m!(1+m)^{r+1}\lambda^{r+1}}\\
~\\
&=\frac{\theta\lambda e}{(e-1)^\theta}\left[e\sum_{m=0}^{\infty}\frac{{(-1)^m}{2^m}}{m!(1+m)^{r+1}} -\sum_{m=0}^{\infty}\frac{(-1)^m}{m!(1+m)^{r+1}}\right]
\end{align*}
By putting r=1, 2, 3... the raw moments can be viewed.
\subsection{\textbf{Moment Generating Function}}
The moment generating function (MGF) of $PGDUSE(\lambda,\theta)$ distribution is given by
\begin{align*}
M_X(t)&=E(e^{tX})\\
~\\
&=\int_0^\infty e^{tx}\frac{\theta\lambda}{(e-1)^\theta}e^{1-\lambda x-e^{-\lambda x}}(e^{1-e^{-\lambda x}}-1)dx\\
~\\
&=\frac{\theta\lambda}{(e-1)^\theta}\int_0^\infty e^{tx-\lambda x-2e^{-\lambda x}}dx-\frac{\theta\lambda}{(e-1)^\theta}\int_0^\infty e^{1+tx-\lambda x-e^{-\lambda x}}dx\\
~\\
&=\frac{\theta\lambda e^2}{(e-1)^\theta}\sum_{m=0}^{\infty}\frac{(-1)^m 2^m}{m!(\lambda+\lambda m-t)}-\frac{\theta\lambda e}{(e-1)^\theta}\sum_{m=0}^{\infty}\frac{(-1)^m}{m!(\lambda+\lambda m-t)}\\
~\\
&=\frac{\theta\lambda e}{(e-1)^\theta}\left[e\sum_{m=0}^{\infty}\frac{{(-1)^m}{2^m}}{m!(\lambda+\lambda m-t)} -\sum_{m=0}^{\infty}\frac{(-1)^m}{m!(\lambda+\lambda m-t)}\right]
\end{align*}

\subsection{\textbf{Characteristic Function}}
The characteristic function (CF) is given by
\begin{equation*}
\phi_X(t)=\frac{\theta\lambda e}{(e-1)^\theta}\left[e\sum_{m=0}^{\infty}\frac{{(-1)^m}{2^m}}{m!(\lambda+\lambda m-it)} -\sum_{m=0}^{\infty}\frac{(-1)^m}{m!(\lambda+\lambda m-it)}\right]
\end{equation*}
where $i=\sqrt{-1}$ is the unit imaginary number.

\subsection{\textbf{Cumulant Generating Function}}
The cumulant generating function (CGF) is given by
\begin{align*}
K_X(t)&=\log \phi_X(t)=\log\left[\frac{\theta\lambda e}{(e-1)^\theta}\left[e\sum_{m=0}^{\infty}\frac{{(-1)^m}{2^m}}{m!(\lambda+\lambda m-it)} -\sum_{m=0}^{\infty}\frac{(-1)^m}{m!(\lambda+\lambda m-it)}\right]\right]\\
~\\
&=\log(\frac{\theta\lambda e}{(e-1)^\theta})+\log\left[e\sum_{m=0}^{\infty}\frac{{(-1)^m}{2^m}}{m!(\lambda+\lambda m-it)} -\sum_{m=0}^{\infty}\frac{(-1)^m}{m!(\lambda+\lambda m-it)}\right]
\end{align*}
where $i=\sqrt{-1}$ is the unit imaginary number.

\subsection{\textbf{Quantile Function}}
The qth quantile $Q(q)$ is the solution of the equation $G(Q(q))=q$.
Hence,
$$Q(q)=\frac{-1}{\lambda}\log(1-\log(q^{\frac{1}{\theta}}(e-1)+1)).$$
The median is obtained by setting $q=0.5$ in the above equation.
Thus,
$$Median=\frac{-1}{\lambda}\log(1-\log(0.5^{\frac{1}{\theta}}(e-1)+1))$$
\subsection{\textbf{Order Statistic}}
Let $X_{(1)},X_{(2)},\ldots,X_{(n)}$ be the order statistics corresponding to the random sample
 $X_1,X_2,\dots,X_n$ of size n from the proposed PGDUSE distribution. 

The pdf and cdf of rth order statistics of the proposed PGDUSE distribution are given by
\begin{equation*}
  g_r(x)=\frac{n!\theta \lambda}{(r-1)!(n-r)!}\frac{e^{1-\lambda x-e^{-\lambda x}}(e^{1-e^{-\lambda x}}-1)^{\theta r-1}}{(e-1)^{2\theta}}\left[1-\left(\frac{e^{1-e^{-\lambda x}}-1}{e-1}\right)^\theta\right]
\end{equation*}
and\\
\begin{equation*}
  G_r(x)=\sum_{i=1}^n\binom{n}{i}\bigg(\frac{e^{1-e^{-\lambda x}}-1}{e-1}\bigg)^{\theta i}\left[1-\bigg(\frac{e^{1-e^{-\lambda x}}-1}{e-1}\bigg)^\theta\right]^{n-i}
\end{equation*}\\
Then, the pdf and cdf of $X_{(1)}$ and $X_{(2)}$ are obtained by
substituting $r = 1$ and $r = n$ respectively in $g_r(x)$ and
$G_r(x)$. It is nothing but, the distribution of Minimum and Maximum
in series and parallel reliability systems, respectively.

\subsection{\textbf{Entropy}}
Entropy quantifies the measure of information or uncertainty.
 An important measure of entropy is R\'{e}nyi entropy. R\'{e}nyi entropy is defined as
$$\gimel_R(\delta)=\frac{1}{1-\delta}\log\bigg(\int g^\delta(x)dx\bigg)$$
where $\delta >0 $ and $\delta\neq 1$.\\
\begin{align*}
\int_0^\infty g^\delta(x)dx&=\frac{(\theta\lambda)^\delta}{(e-1)^{\theta\delta}}\int_0^\infty(e^{\delta-\delta\lambda x-\delta e^{-\lambda x}})(e^{1-e^{-\lambda x}}-1)dx\\
~\\
&=\frac{(e\theta\lambda)^\delta}{(e-1)^{\theta\delta}}\int_0^\infty e^{-\delta\lambda x}e^{-\delta e^{-\lambda x}}\sum_{k=0}^\infty \binom{\delta}{k}(e^{1-e^{-\lambda x}})^{\delta-k}(-1)^k dx\\
~\\
&=\frac{(\theta\lambda e)^\delta}{(e-1)^{\theta\delta}}\int_0^\infty\sum_{k=0}^\infty \binom{\delta}{k}(-1)^ke^{\delta-k}e^{-(\delta-k)\lambda x}e^{-\delta\lambda x}e^{-\delta e^{-\lambda x}}dx\\
~\\
&=\frac{(\theta\lambda e)^\delta}{(e-1)^{\theta\delta}}\sum_{k=0}^\infty \binom{\delta}{k}(-1)^ke^{\delta-k}\sum_{m=0}^\infty(-1)^m\delta^m\int_0^\infty e^{-\lambda mx}e^{-(\delta-k)\lambda x}e^{-\lambda\delta x}dx\\
~\\
&=\frac{(\theta\lambda e)^\delta}{(e-1)^{\theta\delta}}\sum_{k=0}^\infty\sum_{m=0}^\infty \binom{\delta}{k}(-1)^{k+m}\delta^m e^{\delta-k}\int_0^\infty e^{-\lambda(m+2\delta-k)x}dx\\
~\\
&=\frac{(\theta\lambda e)^\delta}{(e-1)^{\theta\delta}}\sum_{k=0}^\infty\sum_{m=0}^\infty \frac{\binom{\delta}{k}(-1)^{k+m}\delta^m e^{\delta-k}}{m+2\delta-k}
\end{align*}\\
The R\'{e}nyi entropy takes the form\\
\begin{align*}
\gimel_R(\delta)&=\frac{1}{1-\delta}\log\left[\frac{(\theta\lambda e)^\delta}{(e-1)^{\theta\delta}}\sum_{k=0}^\infty\sum_{m=0}^\infty \frac{\binom{\delta}{k}(-1)^{k+m}\delta^m e^{\delta-k}}{m+2\delta-k}\right]\\
~\\
&=\frac{\delta}{1-\delta}\log\left[\frac{\theta\lambda e}{(e-1)^\theta}\right]+\frac{1}{1-\delta}\log\left[\sum_{k=0}^\infty\sum_{m=0}^\infty \frac{\binom{\delta}{k}(-1)^{k+m}\delta^m e^{\delta-k}}{m+2\delta-k}\right]
\end{align*}

\section{Estimation}
Here, the estimation of parameters by the method of maximum likelihood is discussed.
 For this, consider a random sample of size n from $PGDUSE(\lambda,\theta)$ distribution. Then the likelihood function is given by,
$$L(x)=\prod_{i=1}^n g(x)=\prod_{i=1}^n \frac{\theta\lambda}{(e-1)^\theta}e^{1-\lambda x_i-e^{-\lambda x_i}}(e^{1-e^{-\lambda x_i}}-1)^{\theta-1}$$
Then the log-likelihood function becomes,
$$\log L=n \log \theta+n\log \lambda-\theta n\log(e-1)-\lambda\sum_{i=1}^{n}x_i+n-\sum_{i=1}^{n}e^{-\lambda x_i}+(\theta-1)\sum_{i=1}^{n}\log(e^{1-e^{-\lambda x_i}}-1)$$
The maximum likelihood estimator(MLE)s are obtained by maximizing
the log-likelihood with respect to the unknown parameters $\lambda$
and $\theta$.
\begin{align*}
  \frac{\partial \log L}{\partial \lambda} &=\frac{n}{\lambda}-\sum_{i=1}^n x_i+\sum_{i=1}^n x_ie^{-\lambda x_i}+(\theta-1)\sum_{i=1}^n\frac{x_ie^{1-\lambda x_i-e^{-\lambda x_i}}}{e^{1-e^{-\lambda x_i}}-1} \\
  \frac{\partial \log L}{\partial \theta} &=\frac{n}{\theta}-n\log(e-1)+\sum_{i=1}^n\log(e^{1-e^{-\lambda
  x_i}}-1).
\end{align*}
These non-linear equations can be numerically solved through
statistical softwares like R with arbitrary initial values.


\section{Data Analysis}
\begin{figure}
  \centering
  \includegraphics[width=5in]{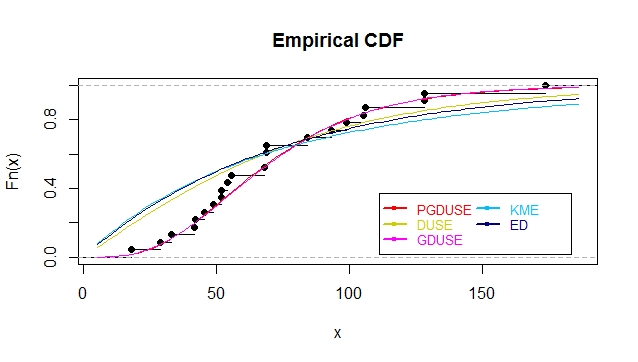}\\
  \caption{\textbf{The empirical cumulative density functions of the models.}}\label{ecdf}
\end{figure}

In this section,  a real data analysis is given to assess how well
the proposed distribution works has been performed.
The data given in Lawless {\cite{LL1982}}  that contains the number of million
revolutions before failure of 23 ball bearings put on life test is considered, see Table \ref{tab1}.\\
\begin{table}[!htb]
    \caption{\textbf{Lawless Data}}\label{tab1}
    \begin{center}
        \begin{tabular}{c c c c c c }
            \hline  \\
            17.88 & 28.92 & 33.00 & 41.52 & 42.12 & 45.60 \\
            48.80 & 51.84 & 51.96 & 54.12 & 55.56 & 67.80\\
            68.64 & 68.64 & 68.88 & 84.12 & 93.12 & 98.64\\
            105.12 & 105.84 & 127.92 & 128.04 & 173.40 &\\
            \hline
        \end{tabular}
    \end{center}
\end{table}
\\
Further, the proposed distribution has been compared with generalized DUS exponential (GDUSE), DUS exponential (DUSE), exponential (ED), and KM exponential distributions. AIC (Akaike Information Criterion), BIC (Bayesian Information Criterion), value of Kolmogorov- Smirnov (KS) statistic, p-value, and log-likelihood value have been used for model selection.
\renewcommand \arraystretch{2}
\begin{table}[!htb]
    \caption{\textbf{MLEs of the parameters, Log-likelihoods, AIC, BIC, K-S Statistics and p-values of the fitted models}}\label{sumtab}
 \begin{center}
    \resizebox{13.5cm}{3.25cm}{
        \begin{tabular}{|c|c|c|c|c|c|c|}
            \hline

            \textbf{Model} & \textbf{MLEs}  & $\mathbf{\log L}$ & \textbf{AIC}  & \textbf{BIC} & \textbf{KS} &   \textbf{p- value} \\

            \hline

            \multirow{2}{*}{\textbf{PGDUSE}} & $\hat{\lambda}$ = 0.03362141& \multirow{2}{*}{-113.003} & \multirow{2}{*}{230.006} & \multirow{2}{*}{232.277} & \multirow{2}{*}{0.11025} & \multirow{2}{*}{ 0.9425} \\
            & $\hat{\theta}$ = 3.80657627 &&&&& \\
            \hline

            \multirow{2}{*}{\textbf{GDUSE}} & $\hat{\alpha}$ = 4.73914452& \multirow{2}{*}{-113.0466} & \multirow{2}{*}{230.0931} & \multirow{2}{*}{232.3641} & \multirow{2}{*}{0.11793} & \multirow{2}{*}{0.9064} \\
            & $\hat{\beta}$ = 0.03553247 &&&&& \\
            \hline

            \textbf{DUSE} & $\hat{a}$ = 0.01824005 & -127.4622 & 256.9244 & 261.1954 & 0.2774 & 0.05804 \\
            \hline

            \textbf{KME} & $\hat{\theta}$ = 0.009544456 & -123.1065 & 248.2129 & 252.4839 & 0.31102 & 0.02337 \\
            \hline

            \textbf{ED} & $\hat{\theta}$ = 0.01384327 & -121.4393 & 244.8786 & 246.0141 & 0.30673 & 0.02639  \\
            \hline

        \end{tabular}
    }
 \end{center}
\end{table}\\
Table \ref{sumtab} elucidates that the proposed distribution gives the lowest AIC, BIC, KS values,
  greatest log-likelihood and p-value. Thus, it can be concluded that the proposed PGDUSE distribution provides a better fit for the given data set when compared with other competing distributions. The empirical cumulative density plot is depicted in Figure \ref{ecdf}.

\section{Conclusion}
\par In this article,  a new class of distribution by generalizing the DUS transformation,
 called the PGDUS transformation is introduced.
  A new lifetime distribution called the PGDUSE distribution with exponential as the baseline distribution is also proposed.
  The generalized form  provides greater flexibility in modelling real datasets.
  Different statistical properties such as  moments, moment generating function, characteristic function, quantile function,
  cumulant generating function, order statistic and entropy are derived. The parameter estimation has been done through
  the method of maximum likelihood. Lastly, a real data analysis is performed to show that the proposed generalization
  can be used effectively to provide better fits.


\end{document}